\documentclass[aps,prc,groupedaddress,showpacs,showkeys]{revtex4}
\usepackage[dvips]{graphicx}
\usepackage{bm}
\usepackage{epsfig}
\usepackage{amsfonts}
\usepackage{amssymb,amscd}
\usepackage{subfigure}
\usepackage{latexsym}
\usepackage{amsmath}
\usepackage{slashed}

\newcommand{\be}{\begin{equation}}
\newcommand{\ee}{\end{equation}}
\newcommand{\bea}{\begin{eqnarray}}
\newcommand{\eea}{\end{eqnarray}}
\newcommand{\bem}{\begin{multline}}
\newcommand{\eem}{\end{multline}}
\newcommand{\beg}{\begin{gather}}
\newcommand{\eeg}{\end{gather}}

\newcommand{\ben}{\begin{eqnarray*}}
\newcommand{\een}{\end{eqnarray*}}

\graphicspath{{Figs/}}
\newcommand{\rr}{\mbox{\boldmath $r$}}

\begin{document}

\title{Intrinsic charm contribution to the prompt atmospheric 
neutrino flux}

\author{A.V. Giannini$^\ddag$,  V.P.  Gon\c{c}alves$^\S$ and 
F.S. Navarra$^\ddag$ }
\affiliation{\ddag\ Instituto de F\'{\i}sica, Universidade de S\~{a}o Paulo\\
C.P. 66318,  05315-970 S\~{a}o Paulo, SP, Brazil}
\affiliation{\S\ Instituto de F\'{\i}sica e Matem\'atica,  Universidade 
Federal de Pelotas\\
Caixa Postal 354, CEP 96010-900, Pelotas, RS, Brazil\\}

\begin{abstract}
In this work we investigate the impact of intrinsic charm on the prompt 
atmospheric neutrino flux. The color  
dipole approach to  heavy quark production is generalized to include  
the contribution of processes initiated by charm quarks. The prompt   
neutrino flux is calculated assuming the presence of  intrinsic charm in the
wave function of the projetile hadron.  
The predictions are compared with previous color dipole results which were   
obtained taking into account only the process initiated by gluons. In   
addition, we estimate the atmospheric (conventional + prompt) neutrino flux  
and compare our predictions with the ICECUBE results for the astrophysical  
neutrino flux. Our results demonstrate that the contribution of the charm 
quark initiated process is non - negligible and that the prompt neutrino 
flux can be enhanced by a factor $\approx$ 2 at large neutrino energies if 
an intrinsic charm component is present in the proton wave function.

\end{abstract}
\maketitle
\vspace{1cm}

\section{Introduction}

Astrophysical neutrinos detected by the IceCube Observatory mark the   
beginning of   neutrino astronomy 
\cite{IceCube_Science,Aartsen:2014gkd,Aartsen:2016xlq}, which allows us 
to  study very high energy  physical processes in the Universe   
\cite{review_icecube,annual_mezsaros}. 
The  atmospheric neutrino flux  is produced in the 
atmosphere  of the Earth (through cosmic-ray interactions with nuclei) and 
it is the main background in studies of  cosmic neutrinos.  
During the last years, several neutrino observatories        
\cite{Abbasi:2010ie,Aartsen:2012uu,Adamson:2012gt,Fukuda:1998ub} have    
studied the  high-energy neutrino flux. The experimental data
indicate that, at low energies ($E_{\nu}\lesssim 10^5$ GeV),  
the measured neutrino flux  is dominated by atmospheric 
neutrinos which come from the decay of light mesons (pions and kaons). This 
flux is called  {\it conventional} atmospheric neutrino flux 
\cite{Honda:2006qj,Barr:2004br,Gaisser:2014pda}. On the other hand, in 
the energy range 10$^{5}$ GeV $< E_{\nu} <$ 
10$^{7}$ GeV, the {\it prompt} atmospheric neutrino flux 
(resulting from the decay of heavy quark hadrons) 
becomes important \cite{ingelman,Martin:2003us,sigl}. The precise knowledge of 
this contribution is crucial for the determination of the cosmic neutrino flux.

The calculation of the prompt atmospheric neutrino flux has been  subject of  
intense activity 
\cite{enb_stasto1,sigl,rojo1,enb_stasto2,rojo2,halzen,laha,prosa,sigl2}.  
Since heavy quarks are an important source of neutrinos, it became 
necessary to describe their production with better accuracy.  
Different treatments (and approximations) of heavy quark production at 
high energies and of the QCD dynamics at small values of 
the Bjorken - $x$ variable were proposed. 
Although the  LHC data on the prompt heavy quark  cross sections       
(see e.g. Refs.~\cite{Aaij:2013mga,Aaij:2015bpa}) helped us to improve  
the description of  heavy meson production at forward rapidities and   
significantly reduced some of the theoretical uncertainties, 
the predictions obtained by different groups can still differ by a 
factor $\ge 2$.
This large uncertainty is due to the fact that the main contribution to  
the prompt neutrino flux comes from  heavy quark production in a  
kinematical range which is not currently probed by the LHC. In  Ref.   
\cite{vicantoni} the authors 
have presented a detailed analysis of the kinematical domains in which   
charm and prompt atmospheric neutrino production from cosmic rays are 
relevant for the IceCube experiment. They 
explored the sensitivity of the corresponding neutrino flux and of the charm 
cross section to the cuts on the maximal $pp$ c.m. energy, to  the 
longitudinal momentum fraction in the target and projectile, to the 
Feynman $x_F$ and to  $p_T$ values included in the calculation.     
They have demonstrated that  in order to address the production of 
high-energy neutrinos ($E_{\nu} >$ 10$^7$ GeV) 
one needs to know the charm production cross section at energies larger  
than those available at the LHC as well as the parton/gluon distributions 
in the region 10$^{-8}$ $< x <$ 10$^{-5}$, which are not presently available 
in collider measurements. Consequently, the presence of new effects,  
which are expected to contribute to small values of $x$ and/or large 
values of the $x_F$, cannot be excluded by current data.

The  production of heavy states  at high energies  is expected to be 
sensitive to  non - linear effects of   QCD dynamics 
\cite{kt_hq,hq_sat,vicmag,cgn,wata,armesto}, which are predicted to be 
enhanced at forward rapidites \cite{cgc}. 
The cross sections at forward rapidities are dominated by collisions of 
projectile partons with large light cone momentum 
fractions ($x_p \rightarrow 1$) with target partons carrying  a very small 
momentum fraction  ($x_t \ll 1$). Consequently, small-$x$ effects coming 
from the non-linear aspects of QCD and from the physics of the 
Color Glass Condensate (CGC)  \cite{cgc} are expected to appear and the  
usual factorization formalism is expected to break down \cite{hq_sat}.  
Recent results \cite{dhj,buw,ptmedio.nois}
indicate that the CGC formalism provides a satisfactory description of 
the experimental data on particle production at forward rapidities.  
Additionally, large - $x$ effects in the projectile are also expected to 
contribute to  heavy quark production at forward rapidities. One of 
the possible new effects  is the presence of intrinsic heavy quarks in 
the hadron wave function (For recent reviews see, e.g. Refs.  
\cite{review_brod_adv,review_brod_prog}). Heavy quarks in the sea of the 
proton can be perturbatively generated by gluon splitting.  Quarks 
generated in this way are usually denoted  {\sl extrinsic} heavy quarks.   
In contrast, the {\sl intrinsic} heavy quarks have multiple connections to     
the valence quarks of the proton and thus are sensitive to its nonperturbative  
structure. 

The existence of the intrinsic charm (IC) component was first   
proposed long ago in Ref. \cite{bhps} (see also Ref. \cite{hal}) and since   
then other models of IC  have been  discussed \cite{pnndb,wally99}. Its 
existence implies a large enhancement of the charm distribution     
at large  $x$  ($> 0.1$) in comparison to the extrinsic charm prediction.    
Moreover, due to the momentum sum rule, the gluon distribution is also  
modified by the inclusion of  intrinsic charm. The  
intrinsic charm (IC) component of the  proton wave function was 
included in recent global analyses,  performed by different theoretical 
groups, in order to constrain the parton distributions. 
Although the resulting IC distributions are compatible with the world data, 
the amount of IC  in the proton wave function  is still  subject of intense 
debate  \cite{wally16,brod_letter} and has motivated a large number of 
phenomenological studies  (See e.g. Refs. \cite{russos,bailas,biw16}).

One of the most  direct consequences of the intrinsic charm component is that  
it gives rise to heavy mesons with large fractional momenta relative to  
the beam particles,  affecting  the $x_F$ and rapidity 
distribution of charmed particles. This aspect was explored e.g. in  
Refs. \cite{ingelman,kniehl,npanos,prdnos}. In particular, 
in Ref. \cite{prdnos}  we studied  $D$ - meson production at forward rapidities taking into account  non - linear effects in the QCD dynamics and the intrinsic charm (IC) component  
of the proton wave function.  The results show that, at the LHC, the intrinsic charm 
component  changes  the $D$ rapidity distributions  in a region  which is beyond the
coverage of the LHCb detector.  At higher energies, as those probed in  neutrino 
observatories, our results indicated that the IC component dominates the rapidity 
and $x_F$ distributions, with the latter being  enhanced by a factor 6 -- 8 in the 
$0.2 \le x_F \le 0.8$ range.  In Ref. \cite{prdnos} we have pointed out that one of 
the basic consequences of this enhancement is the modification of the prompt neutrino 
flux at large energies. The main purpose of this work is to estimate the impact of  
the intrinsic charm component on the prompt neutrino flux taking into account   
non - linear effects in the QCD dynamics. It is important to emphasize that the 
contribution associated to a heavy quark in the initial state was not taken into account 
in previous studies using the color dipole formalism performed e.g. in Refs. 
\cite{ers,enb_stasto2}. The inclusion of heavy quarks in the initial state in the color 
dipole approach  is done here for the first time. Additionally, previous estimates 
of the intrinsic charm contribution to the neutrino flux \cite{laha,halzen} 
have made use of 
phenomenological models of the 
$x_F$ - distribution in  $\Lambda_c$ and $D$ production,     
constrained only by the scarce low energy experimental data  
and then extrapolated 
 to higher 
energies.  In contrast, in our analysis the basic ingredients of the calculations, 
the charm/gluons PDF's and the dipole - hadron  amplitude, have been constrained by the most 
recent LHC data and  non - linear effects in the QCD dynamics are taken into account.

This paper is organized as follows. In the next Section we present a brief review of the formalism to calculate the charm production at forward rapidities, as well as we discuss the impact of an intrinsic charm component in the Feynman $x_F$ - distribution. In Section \ref{res} we present our results for the prompt neutrino flux. The contribution of the different components is analyzed and the impact of the intrinsic component is estimated. In particular, the predictions for the atmospheric (conventional + prompt) neutrino flux is compared with the recent ICECUBE results for the flux of the astrophysical neutrinos. Finally, in Section \ref{conc} we summarize our main results and conclusions.

\section{Formalism}
\label{form}

In the analysis performed in this paper, we will closely follow the  
procedure described  in detail in Refs. \cite{vicantoni,prdnos}.  
In what follows we will only 
present the main aspects of the formalism and refer the reader to Refs. \cite{vicantoni,prdnos} 
for more details. As in Ref. \cite{vicantoni}, we will calculate the prompt neutrino flux  
using the semi-analytical $Z$-moment approach, proposed many years ago in Ref.~\cite{ingelman} and
discussed in detail e.g.~in Refs.~\cite{ers,rojo1}.
In this approach, a set of coupled cascade 
equations for the nucleons, heavy mesons and leptons (and their antiparticles) 
fluxes is solved, with the equations being expressed in terms of the nucleon-to-hadron
($Z_{NH}$), nucleon-to-nucleon ($Z_{NN}$), hadron-to-hadron  ($Z_{HH}$) 
and hadron-to-neutrino ($Z_{H\nu}$) $Z$-moments. 
These moments are inputs in the calculation of the prompt neutrino flux
associated with the production of a heavy hadron $H$ and its decay into a
neutrino $\nu$ in the low- and high-energy regimes. We will focus on 
vertical fluxes  and will 
assume that the cosmic ray flux $\phi_N$  can be described by a broken 
power-law spectrum \cite{prs} or by the H3a spectrum proposed in Ref. \cite{gaisser}, with the incident flux being
represented by protons. Moreover, we will assume that the charmed 
hadron $Z$-moments can be expressed in terms of the charm $Z$-moment 
as follows: $Z_{pH} = f_H \times Z_{pc}$, where $f_H$ is the fraction of charmed
particle which emerges as a hadron $H$. As in Ref.~\cite{ers}, 
we will assume that $f_{D^0} = 0.565$, $f_{D^+} = 0.246$, 
$f_{D_s^+} = 0.080$ and $f_{\Lambda_c} = 0.094$.

The charm $Z$-moment at high energies can be expressed by 
\begin{eqnarray}
Z_{pc} (E) =  \int_0^1 \frac{dx_F}{x_F} \frac{\phi_p(E/x_F)}{\phi_p(E)} 
\frac{1}{\sigma_{pA}(E)} \frac{d\sigma_{pA \rightarrow charm}(E/x_F)}{dx_F} \,\,,
\label{eq:zpc}
\end{eqnarray}
where $E$ is the energy of the produced particle (charm), $x_F$ 
is the Feynman variable, $\sigma_{pA}$ is the inelastic proton-Air
cross section, which we assume to be given as in Ref.~\cite{sigl}, and 
$d\sigma/dx_F$ is the differential cross section for the charm production, 
which we assume to be given by $d\sigma_{pA \rightarrow charm}/dx_F = 2
\, d\sigma_{pA \rightarrow c \bar{c}}/dx_F$. 
One of the main inputs in the $Z$-moment  approach is the Feynman $x_F$ 
distribution of the heavy quarks produced in hadronic collisions. 
As discussed in Ref. \cite{vicantoni}, the main 
contribution to the prompt neutrino flux comes from large values of $x_F$, that are associated with  
heavy quark production at forward rapidities. Moreover, as the production of neutrinos at a given neutrino 
energy, $E_{\nu}$, is determined by collisions of cosmic rays with nuclei in
the atmosphere at energies that are a factor of order 100-1000 larger,   the prompt neutrino flux measured 
in the kinematical range that is probed by the IceCube Observatory and future neutrino telescopes is directly 
associated with the treatment of the heavy quark production cross section at high energies. Following 
Ref. \cite{prdnos}, we will use the approaches developed in Refs. \cite{nnz,boris} and \cite{npanos} for the 
treatment of  heavy quark production induced by gluon - gluon and charm - gluon interactions, respectively,  
and we will take into account  non - linear effects in the QCD dynamics. The contribution of the $q \bar{q} \rightarrow c \bar{c}$  subprocess will be disregarded, since it is negligible at the energies and rapidities relevant for the 
prompt neutrino flux \cite{vicantoni}.  

\begin{figure}[t]
 \begin{center}
 \begin{tabular}{ccc}
 \includegraphics[width=0.3\textwidth]{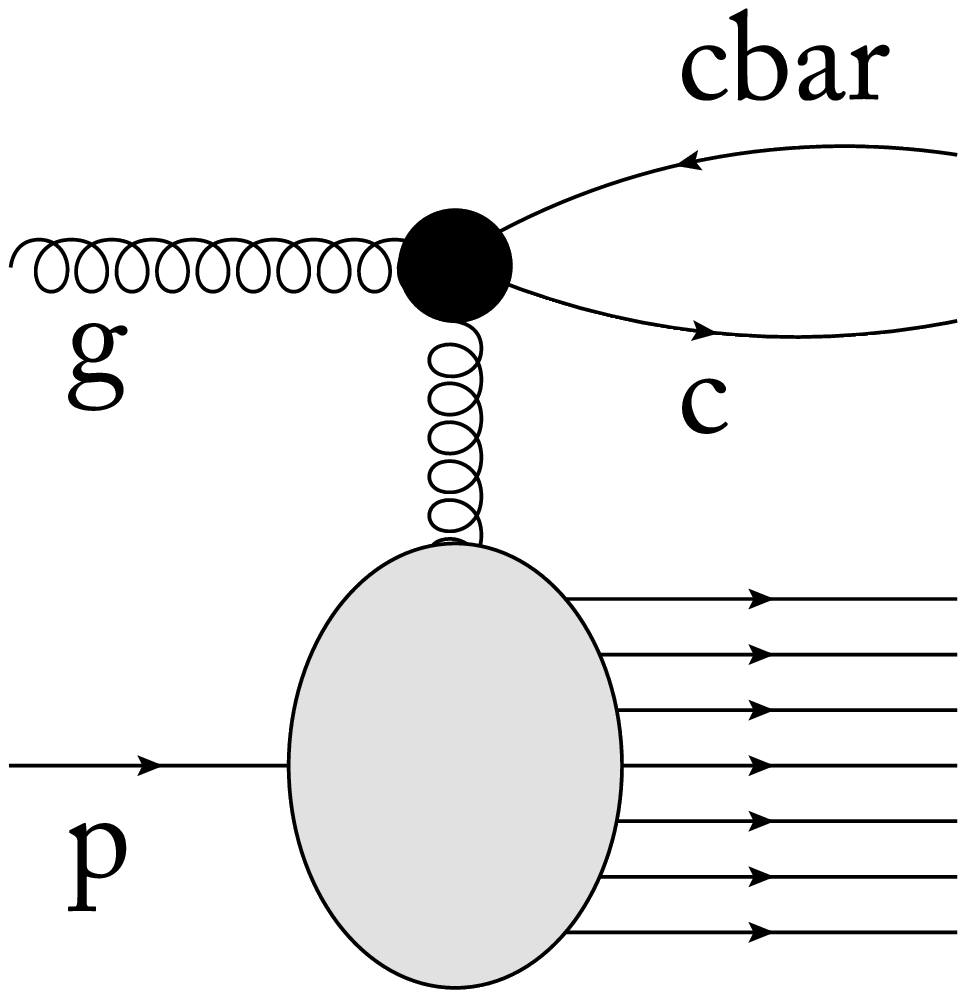} & \,\,\,\,\,\hspace{0.5cm}\,&
 \includegraphics[width=0.3\textwidth]{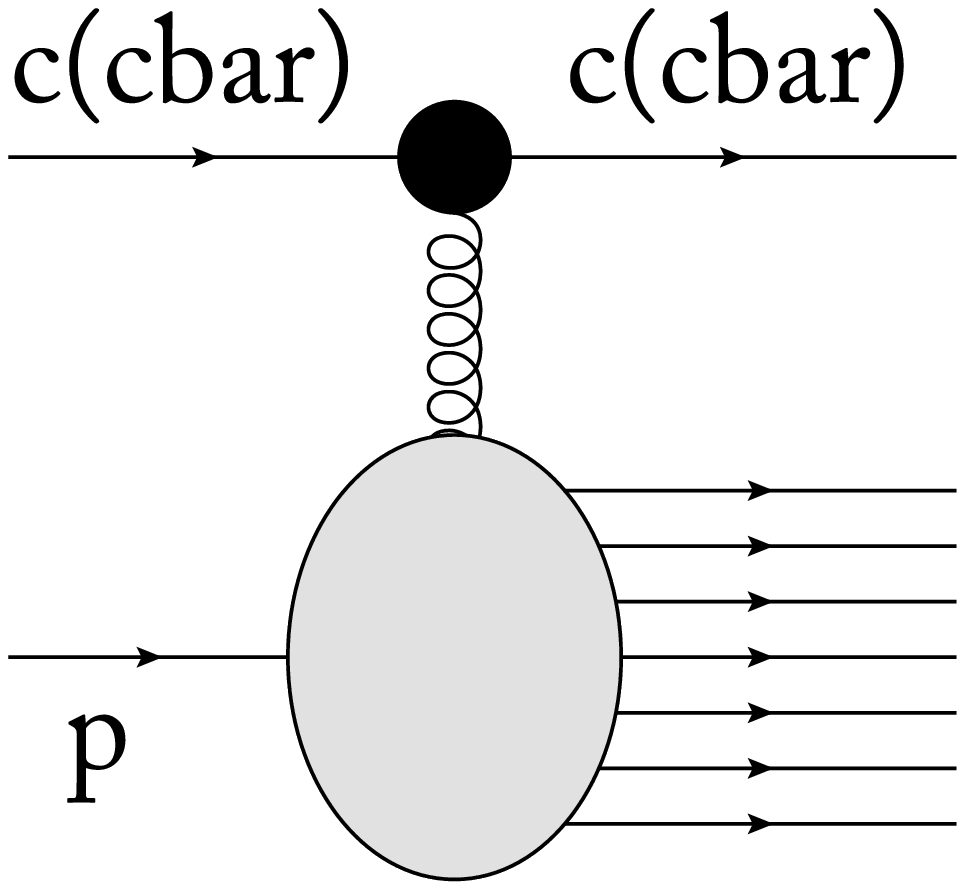} \\
 (a) & \,\,& (b)
 \end{tabular}
  \end{center}
 \caption{Contributions to  charm production at high energies and forward rapidities. (a)  Contribution from gluon - initiated processes. (b) 
Contribution from charm in the initial state.}
  \label{Fig:diagrama}
\end{figure}

The basic idea  in Ref. \cite{prdnos} 
is that at forward rapidities, the projectile (dilute system) evolves according to the linear DGLAP dynamics and  
the target (dense system) is treated using the CGC formalism. 
In this approach, the charm $x_F$ - distribution will be determined by the contribution of the two diagrams presented in Fig. \ref{Fig:diagrama}, which are associated to the 
gluon and charm - initiated processes. The first contribution, represented in Fig. \ref{Fig:diagrama} (a),  can be estimated using the color dipole picture \cite{nnz,boris}, which implies that the rapidity distribution for the  charm production in a  $h_1 h_2$ collision can be expressed as follows
\begin{equation}
\frac{d\sigma}{dy} =  x_1g^{h_1}(x_1,\mu_F^2) \,
\sigma (gh_2 \rightarrow \{c\bar{c} \}X)\,\,,
\label{dsdy_gluons}
\end{equation}
where $x_1g_{p}(x_1,\mu _F)$ is the projectile gluon distribution, the cross section  $\sigma (gh_2 \rightarrow \{c\bar{c}\} X)$ describes  
charm production in a gluon - nucleon interaction, $y$  is the rapidity of the pair and $\mu_F$ is the 
factorization  scale.  Moreover,  the cross section of the process  $g + h_2 \rightarrow c \bar{c} X$ is given by:
\begin{equation}
\sigma(g h_2 \rightarrow\{c\bar{c}\}X) = \int _0^1 d \alpha \int d^2\rr \,\,
\vert \Psi _{g\rightarrow c\bar{c}} (\alpha,\rr)\vert ^2
\,\, \sigma^{h_2} _{c\bar{c}g}(\alpha , \rr)
\label{sec1}
\end{equation}
where  $\alpha$ ($\bar{\alpha} \equiv 1 - \alpha$) is the longitudinal momentum fraction carried by the quark (antiquark), ${\rr}$ is the transverse 
separation of the pair,   $\Psi _{g\rightarrow c\bar{c}}$ is the 
light-cone (LC) wave function of the 
transition $g \rightarrow  c \bar{c} $ (which is calculable perturbatively and is proportional to $\alpha_s$)  and $  \sigma^{h_2}_{c\bar{c}g}$  is the 
scattering cross section of a color neutral quark-antiquark-gluon system on the 
hadron target $h_2$ \cite{nnz,boris}. The three - body cross section is given in terms of the dipole - nucleon cross section $\sigma _{c\bar{c}}$ as follows:
\begin{equation}
\sigma^{h_2}_{c\bar{c}g}(\alpha , \rr) = \frac{9}{8}[\sigma _{c\bar{c}}(\alpha \rr) + \sigma _{c\bar{c}}(\bar{\alpha} \rr)]
- \frac{1}{8}\sigma _{c\bar{c}}(\rr)\,\,.
\label{sec2}
\end{equation}
Finally, the dipole - nucleon cross section can be expressed in terms of the forward scattering amplitude ${\cal{N}} (x,\rr)$, which is determined by the QCD dynamics, as follows:
\begin{equation}
\sigma_{c \bar c}(x,\rr) = \sigma_{0} {\cal N}(x,\rr)
\label{sigzero}
\end{equation} 
where $\sigma_0$ is a free parameter usually determined by a fit of the HERA data. 
On the other hand, the contribution of the charm initiated process, represented in Fig. \ref{Fig:diagrama} (b), is given by \cite{npanos}
\begin{eqnarray}
{d\sigma \over dy} = 
{1 \over (2\pi)^2} \int d^2p_T
f_{c/h_1}(x_1,\mu_F^2)\,\, \sigma_0 \, \widetilde{\cal N} \left(x_2, {p_T}\right)
\label{dNdy_quarks}\,.
\end{eqnarray}
where   $x_{1,2}$ is defined by $x_{1,2} = p_{T} e^{\pm y}/\sqrt{s}$,  $f_{c/h_1}$ represents the projectile charm distribution and $\widetilde{\cal N} (x, {p_T})$ is the Fourier transform of the scattering amplitude ${\cal N}(x,\rr)$.


In order to estimate the contributions of the gluon and charm - initiated processes, described by Eqs. (\ref{dsdy_gluons}) and (\ref{dNdy_quarks}),  we should to assume a model to describe  the dipole - nucleon  scattering amplitude ${\cal N}(x_2,\rr)$, as well as a parametrization for gluon and charm distributions in the projectile. The the dipole - nucleon  scattering amplitude ${\cal N}(x_2,\rr)$ 
describes the interaction of a color dipole of size $r$ with the nucleon and involves the QCD dynamics 
at high energies. Such quantity contains all the information  about the initial state of the hadronic wavefunction and therefore 
about the non-linearities and quantum effects which are characteristic of a system such as the  CGC (For reviews, see 
e.g. \cite{cgc}). Formally its evolution is usually described in the mean field approximation of the CGC formalism by 
the BK equation \cite{bk}. Its analytical solution is known only in some special cases.  Advances have been made in 
solving the BK equation numerically 
\cite{la11}. Since the BK equation still lacks a formal solution in all kinematical space, several groups have 
constructed phenomenological models for the dipole 
scattering amplitude. These models have been used to fit the RHIC, LHC and HERA data \cite{dhj,buw,dips,IIM,Soyez}. 
As in Ref. \cite{prdnos}, in our analysis we will use the BUW model for ${\cal N}$,  originally proposed in 
Ref. \cite{buw}, which assumes  that
${\cal{N}}$ can  be modelled  through a simple Glauber-like formula,
\begin{eqnarray}
{\cal{N}}(x,\rr) = 1 - \exp\left[ -\frac{1}{4} (\rr^2 Q_s^2(x))^{\gamma (x,\rr^2)} \right] \,\,,
\label{ngeral}
\end{eqnarray}
where $Q_s(x)$ is the saturation scale and $\gamma$ is the anomalous dimension of the target gluon distribution. The speed with which we move from the non-linear 
regime to the extended geometric scaling regime and then from the latter to the linear regime is what differs the BUW from other phenomenological models. This transition 
speed is dictated by the behavior of the anomalous dimension $\gamma (x,\rr^2)$, which is assumed in  the BUW \cite{buw} dipole model to be given by
\begin{equation}
\label{dip_adimension}
\gamma(\omega = p_{T}/Q_{s})_{BUW} = \gamma_{s} + (1-\gamma_s)\frac{(\omega^a-1)}{(\omega^a-1)+b}
\end{equation}
where $a$, $b$ and $\gamma_s$ are free parameters to be fixed by fitting  experimental data. In Ref. \cite{ptmedio.nois},  the original parameters of the 
BUW model  were updated in order to  make this model compatible with all existing data.  In particular, the recent LHC data on light hadron production at forward 
rapidity are satisfactorily reproduced by the updated model.

\begin{figure}[t]
 \begin{center}
 \begin{tabular}{ccc}
 \includegraphics[width=0.45\textwidth]{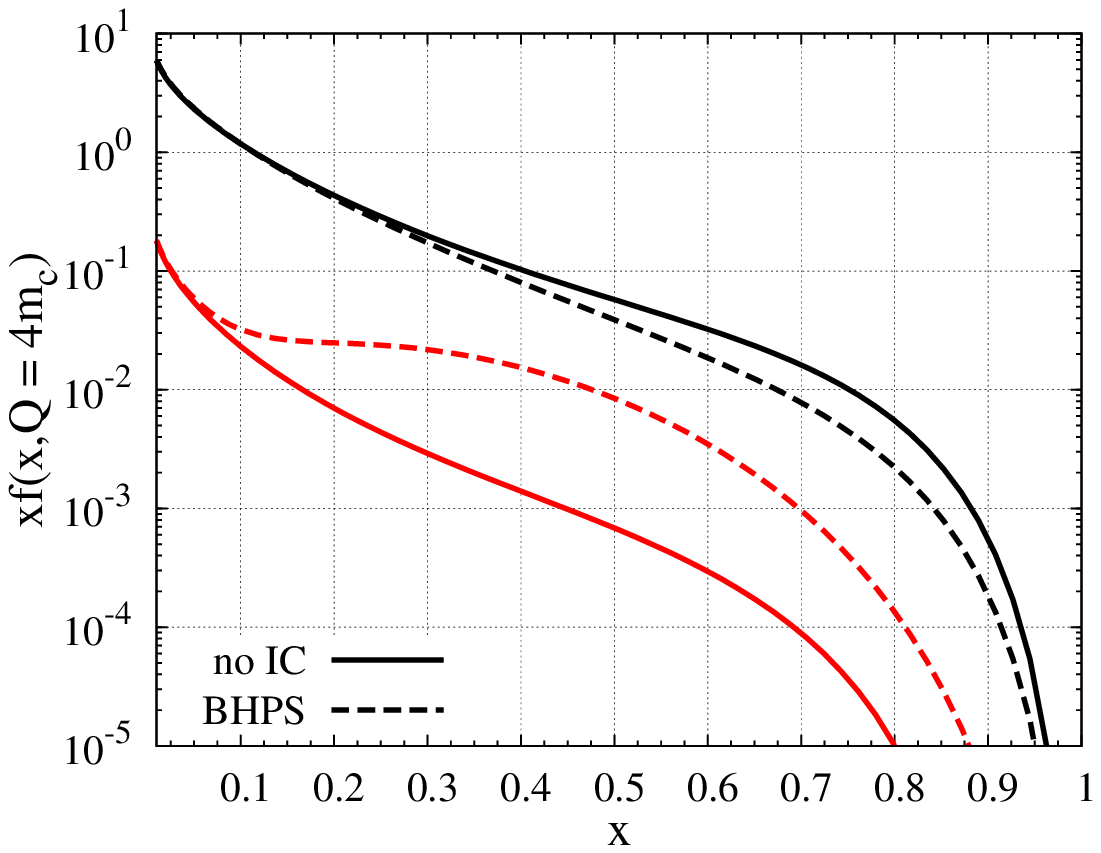} & \,\,\,\,\,\hspace{0.5cm}\,&
 \includegraphics[width=0.45\textwidth]{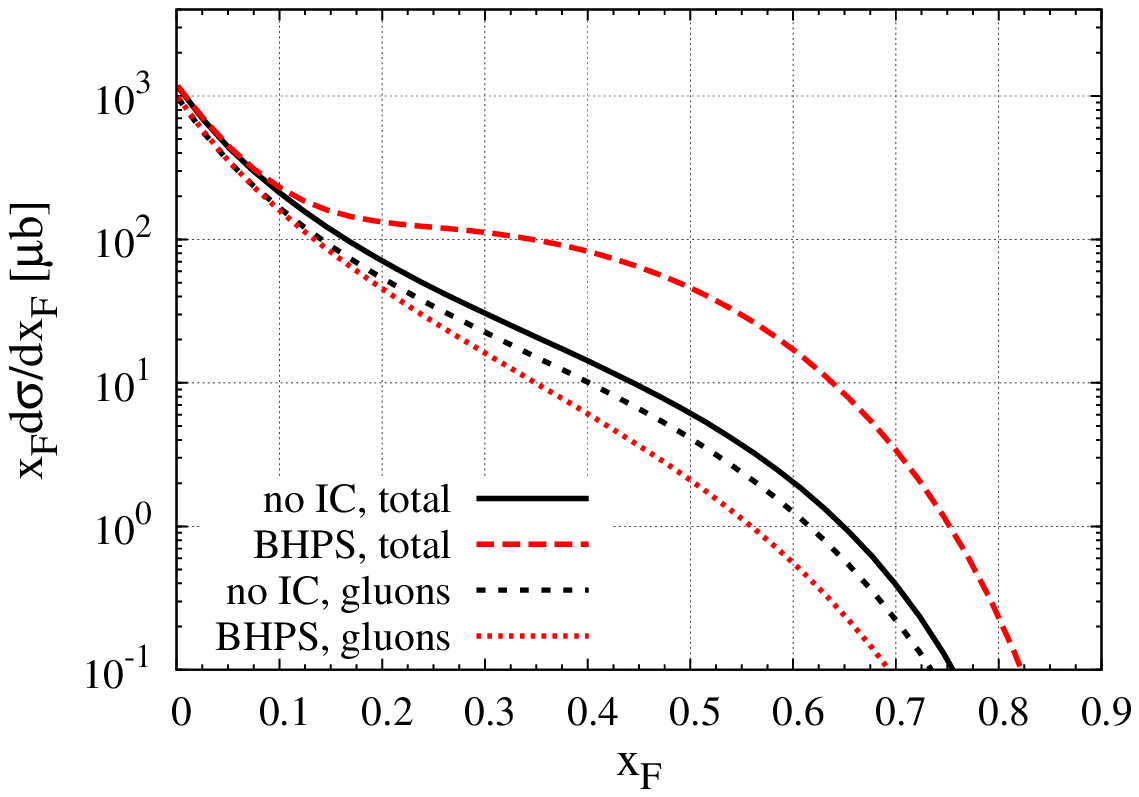} \\
 (a) & \,\,& (b)
 \end{tabular}
  \end{center}
 \caption{(a) Comparison between the BHPS and No IC predictions for the charm (lower red curves) and gluon (upper black curves) distributions. (b) Comparison between the BHPS and No IC predictions for the Feynman $x_F$ distributions considering $pp$ collisions at $\sqrt{s} = 13$ TeV.}
  \label{Fig:intrinsic}
\end{figure}

In order to quantify the impact of the intrinsic charm component on the projectile wave function, we will use in our calculations the 
next - to - leading order  CTEQ 6.5 parametrization for the  parton distributions \cite{cteq}. In Ref. \cite{cteq}, the CTEQ group 
determined the shape and normalization of the IC distribution in the same way 
as they do for other parton species. In fact, they find several IC distributions 
which were compatible with the world data. In what follows we will only consider the parametrization based on the BHPS model \cite{bhps}, obtained assuming that the average longitudinal momentum fraction carried by the charm and anticharm is $\langle x_{c\bar{c}}\rangle = 2\%$ at the initial scale of the QCD evolution.
The BHPS model assumes 
that the nucleon light cone wave function has  higher Fock states,  one of them being
$|q q q c \overline{c}>$. The probability of finding the nucleon in this
configuration is given by the inverse of the squared invariant mass of the
system. Because of the heavy charm mass, this probability as a function of the
quark fractional momentum, $P(x)$, is very hard, as compared to the one obtained
through the DGLAP evolution. Such assumptions are used to describe the shape of the parametrization at the initial scale of the DGLAP evolution, which is evolved for larger scales in order to contrain, by a global data fit procedure, the parton distributions of the proton. In Fig. \ref{Fig:intrinsic} (a) we compare the predictions of the BHPS model with those obtained disregarding the presence of an intrinsic component (denoted No-IC hereafter). In the case of the charm distribution (lower red curves), the BHPS model predict a large enhancement of the distribution 
at large  $x$  ($> 0.1$). In Fig. \ref{Fig:intrinsic} (a) we also present the corresponding gluon distributions (upper black curves). Due to the 
momentum sum rule, the gluon distribution is also  modified by the inclusion of  intrinsic charm. In particular, the BHPS  model imply a 
suppression in the gluon distribution at large  $x$. The impact of these different models on the Feynman $x_F$ distribution for the charm production in $pp$ collisions at $\sqrt{s} = 13$ TeV is presented in Fig. \ref{Fig:intrinsic} (b). 
We  present the sum of the gluon and charm contributions, denoted ``total'' in the figure, as well present separately the gluon contribution. We have that in the No IC case, the distribution is dominated by the gluon initiated process. In contrast, when  intrinsic charm is included, the behavior of the distribution in the intermediate $x_F$ 
range ($0.2 \le x_F \le 0.8$) is strongly modified.   
It is important to emphasize that the contribution of the  gluonic component decreases in this kinematical range, as expected from the analysis of the parton distributions presented in Fig. \ref{Fig:intrinsic} (a). As we will show in the next Section, these modifications in the distribution, associated the presence of an intrinsic component,  has important implications in the prompt neutrino flux.

Before to present our predictions for the prompt neutrino flux in the next Section, two  comments are in order. In our analysis we will consider the next - to - leading order  CTEQ 6.5C parametrization for the  parton distributions \cite{cteq}, which is provided by the CTEQ group. It is important to emphasize that the CTEQ-TEA group  has also performed a global analysis of the recent experimental data including an intrinsic 
charm component, which is available in the CT14 parametrization \cite{cteq14}. However, this analysis has been performed at next-to-next-to-leading order. As demonstrated in Ref. \cite{rauf}, the  predictions of the dipole approach for the charm production agree with those obtained using the collinear formalism at NLO, in the kinematical range where the equivalence between the approaches is expected. Therefore,  we believe that it is more consistent to use in our calculations PDFs obtained at NLO. Second, the CTEQ 6.5C parametrization is used in this paper for consistence with our previous study \cite{prdnos}, where we have performed a detailed analysis of the $D$ - meson production at the LHC energies considering different models for the intrinsic charm component. In particular, in Ref. \cite{prdnos} we have compared the BHPS predictions with those obtained using the Meson Cloud model \cite{pnndb,wally99}. Such model is not considered in the CTEQ 6.6C parametrization, which is an improved version of CTEQ 6.5C one.  We have verified that the modifications in our predictions for the neutrino flux are negligible if the CTEQ 6.6C parametrization is used as input in our calculations.

\section{Results}
\label{res}

In what follows we will present our estimates for the prompt atmospheric neutrino flux using the ingredients discussed 
above.   In our analysis, we will assume $m_c = 1.5$ GeV and $\mu^2 = 4 m_c^2$ and  compare the No IC predictions with the BHPS one. Moreover, in order to estimate the contribution associated to  heavy quarks in the initial state, we also will compare our full predictions, estimated taking into account 
gluons and quarks in the initial state, with those derived disregarding the charm contribution, as was done in previous 
studies using the color dipole approach.

\begin{figure}[t]
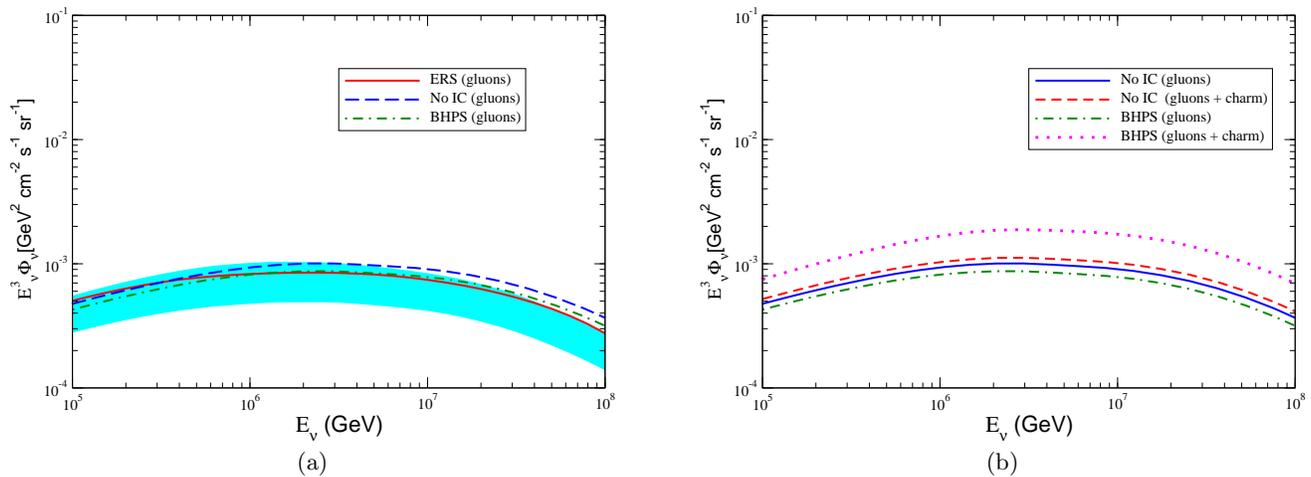

 \begin{center}
 \begin{tabular}{ccc}
 \includegraphics[width=0.45\textwidth]{comps2_withenberg.eps} & \,\,\,\,\,\hspace{0.5cm}\,&
 \includegraphics[width=0.45\textwidth]{comps2.eps} \\
 (a) & \,\,& (b)
 \end{tabular}
  \end{center}
 \caption{Energy dependence of the prompt neutrino flux, normalized by a factor $E_{\nu}^3$. 
(a) Comparison of our predictions with  the results obtained in Refs. \cite{ers,enb_stasto2} 
using the color dipole approach. The results derived in \cite{enb_stasto2} considering different 
phenomenological saturation models for the dipole - proton scattering amplitude are represented by the 
cyan band.  (b)  Comparison of the No IC and BHPS predictions considering the gluon and charm   
contributions and only the gluon one in the calculation of the  $x_F$ distribution.}
  \label{Fig:fluxE3}
\end{figure}

Initially, let's consider that the primary proton flux is described by a broken power-law spectrum. Our motivation to assume this model is associated to the fact that previous estimates for the prompt neutrino flux \cite{ers,enb_stasto2}, which we would like compare our results, has been obtained using this model.
In Fig. \ref{Fig:fluxE3} we present our predictions for the energy dependence of the prompt atmospheric neutrino flux, 
normalized by a factor $E_{\nu}^3$. In the left panel, we compare our predictions with those obtained in Refs.
 \cite{ers,enb_stasto2} using the color dipole approach and different phenomenological models based on saturation 
physics. The predictions obtained in Ref. \cite{ers} are denoted by ERS and those derived in Ref. \cite{enb_stasto2} 
by the cyan band.  As in  Refs. \cite{ers,enb_stasto2} only the processes initiated by gluon were taken into account, 
here, for the sake of comparison, we present our predictions derived considering only this channel as well. At low 
energies ($E_{\nu} \le 10^6$ GeV),
the ERS and No IC predictions are very similar. On the other hand, at large energies  
($E_{\nu} \ge 10^7$ GeV), the No IC predictions imply a larger prompt neutrino flux than that derived in Refs. 
\cite{ers,enb_stasto2}. Such behavior can be attributed to the 
model of the dipole - proton scattering amplitude 
used in our calculations, which differs from those used in  Refs. \cite{ers,enb_stasto2}. As demonstrated e.g. in 
Ref. \cite{betemps}, the BUW model predicts a slower transition between the linear and non - linear regimes of the QCD 
dynamics than e.g. the IIM-S one \cite{IIM,Soyez} used in Refs. \cite{ers,enb_stasto2}. Consequently, the BUW model 
implies a faster growth of the heavy quark cross section with the energy, which has  direct impact on the prompt neutrino 
flux at high energies.   On the other hand, if the gluon distribution associated to the BHPS parametrization is used as 
input in the calculations, we find that the resulting predictions are suppressed in comparison to the No IC one. As discussed 
before, in the  BHPS parametrization the IC component 
is taken into account and this implies that 
a larger amount of the proton momentum is carried by the charm quarks. Due to the momentum sum rule, the amount carried by the 
other partons, in particular the gluons, will be reduced. Therefore, the gluon distributions associated to the parton 
parametrizations  which include  an intrinsic component are, in general, smaller than those derived disregarding this 
component [See Fig. \ref{Fig:intrinsic} (a)]. Such reduction 
explains the result observed in the left panel of 
Fig. \ref{Fig:fluxE3}.

In Fig. \ref{Fig:fluxE3} (b) we estimate the impact of the charm initiated processes on the energy dependence 
of the prompt neutrino flux. As expected, the inclusion of this new channel of charm production leads to an enhancement of the 
flux. The magnitude of this enhancement depends on the details of the model of the charm distribution. When the  intrinsic charm 
component is disregarded [the No IC (gluons + charm) curve in Fig. \ref{Fig:fluxE3}]  
the impact of the charm initiated processes is small. Such result is expected, since the magnitude of the extrinsic charm 
distribution for small values of $\mu^2$ is small. On the other hand, if the intrinsic component is taken into account, we 
have a large enhancement of the prompt flux.

\begin{figure}[t]
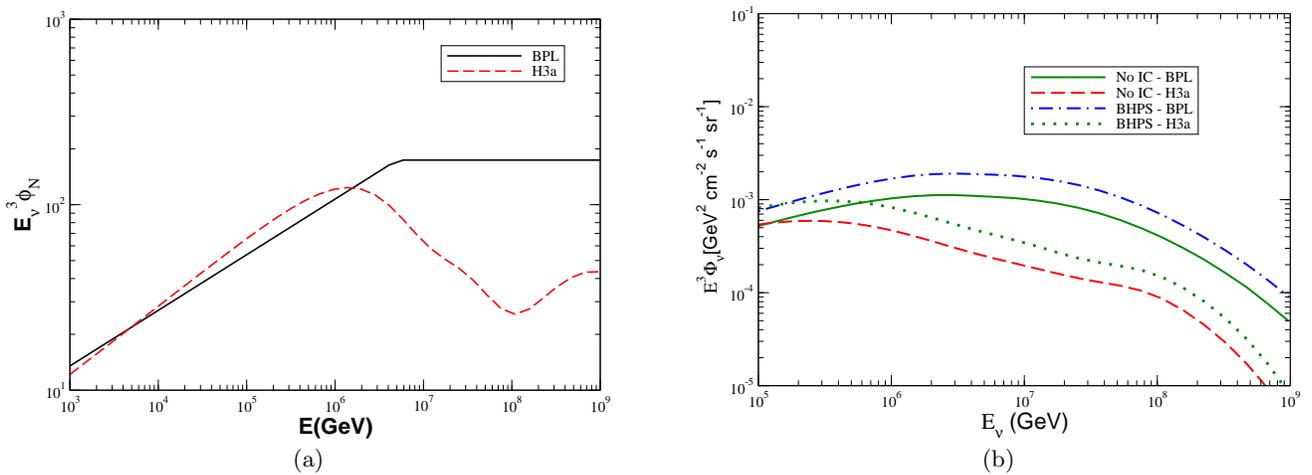

 \begin{center}
 \begin{tabular}{ccc}
 \includegraphics[width=0.45\textwidth]{comp_flux_primarios.eps} & \,\,\,\,\,\hspace{0.5cm}\,&
 \includegraphics[width=0.45\textwidth]{compIC_fluxes.eps} \\
 (a) & \,\,& (b)
 \end{tabular}
  \end{center}
 \caption{(a) Comparison between the BPL and H3a models for the  all - nucleon primary cosmic ray spectra. (b) Energy dependence of the prompt neutrino flux, normalized by a factor $E_{\nu}^3$, obtained using the BPL and H3a models as input in the calculations. }
  \label{Fig:h3a}
\end{figure}

One important question is the dependence of our results on the model assumed for the primary nucleon spectrum.  Although the broken power - law (BPL) approximation of the cosmic ray nucleon flux has been used to evaluate the prompt neutrino flux in almost all early references, more recent publications also present the predictions obtained considering a modern description of the primary spectrum proposed by Gaisser in Ref. \cite{gaisser}. In what follows we also will estimate the prompt flux using the H3a spectrum, which assumes that the spectrum is given by a composition of  3 populations and 5 representative nuclei, with the set of parameters determined by a global fit of the cosmic ray data \cite{gaisser}. A comparison between the BPL and H3a spectra is performed in Fig. \ref{Fig:h3a} (a). We have that for primary energies in the range $10^4 \lesssim E \lesssim 10^6$ GeV, the H3a spectrum is larger than the BPL one. On the other hand, for $E \gtrsim 10^6$ GeV, the H3a spectrum is smaller than the BPL one, and a structure associated to its composition is present. In Fig. \ref{Fig:h3a} (b) we present our predictions for the energy dependence of the prompt neutrino flux obtained using the BPL and H3a models for the primary nucleon flux. As expected from Fig. \ref{Fig:h3a} (a), the H3a predictions  are smaller than the BPL one at large neutrino energies. However, our results indicate a large impact of the intrinsic charm also is present if the H3a spectrum is considered.

\begin{figure}[t]
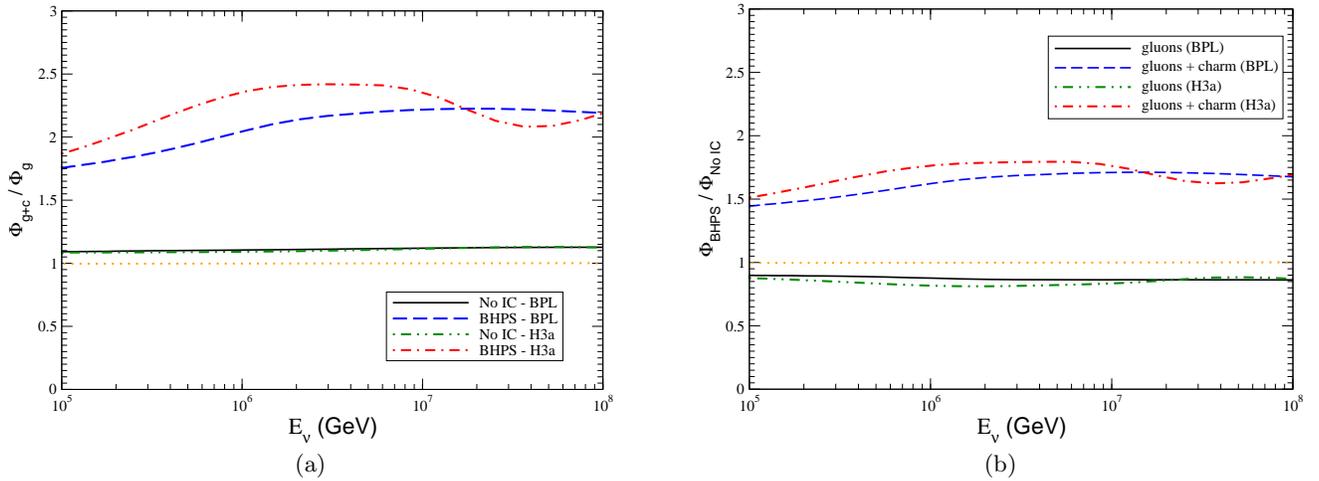

 \begin{center}
 \begin{tabular}{ccc}
 \includegraphics[width=0.45\textwidth]{ratiogluonquark2.eps} & \,\,\,\,\,\hspace{0.5cm}\,&
 \includegraphics[width=0.45\textwidth]{ratioBHPS2.eps} \\
 (a) & \,\,& (b)
 \end{tabular}
  \end{center}
 \caption{Energy dependence of the ratios between prompt neutrino fluxes. (a) Ratio between the flux calculated considering the gluon and 
charm channels ($\Phi_{g+c}$) and that derived assuming only the gluon 
channel ($\Phi_{g}$). (b) Ratio between the flux derived assuming 
the presence of this component  ($\Phi_{BHPS}$) with that calculated 
disregarding this component ($\Phi_{No IC}$). }
  \label{Fig:ratios}
\end{figure}

A more detailed estimate of the contribution of the charm initiated process and the  intrinsic charm component is obtained by the analysis of Fig.  \ref{Fig:ratios}. In Fig. \ref{Fig:ratios} (a)  we  present our results for the ratio between the flux calculated considering the gluon and charm channels ($\Phi_{g+c}$) and that derived assuming only the gluon channel ($\Phi_{g}$). The results associated to the BPL and H3a models are presented for comparison.  When the IC component is absent, the impact 
of the charm channel is $\approx 10 \%$ and almost energy independent. In contrast, if an IC component is present, the impact 
increases with the neutrino energy and becomes larger than $200 \%$ at large energies, with the H3a predictions being slightly larger than BPL one for $E_{\nu} \lesssim 10^7$ GeV. Another way to estimate the impact of 
the IC component is to calculate  the ratio between the flux derived assuming the presence of this component  
($\Phi_{BHPS}$) and the flux calculated disregarding this component ($\Phi_{No IC}$). The results are presented in Fig. 
\ref{Fig:ratios} (b).  They  indicate that if only the gluon channel is considered, the presence of the IC 
component implies a reduction of the prompt neutrino flux of $\approx 10 \%$. On the other hand, when the charm channel is 
included, we predict an enhancement of $\approx 160 \%$ in the flux. These 
results strongly suggest that a generalized treatment of 
heavy quark production, taking into account  charm initiated processes, is crucial to obtain realistic predictions
of the  prompt flux using the color dipole approach, especially if an intrinsic charm component is present in the proton wave function.

\begin{figure}[t]
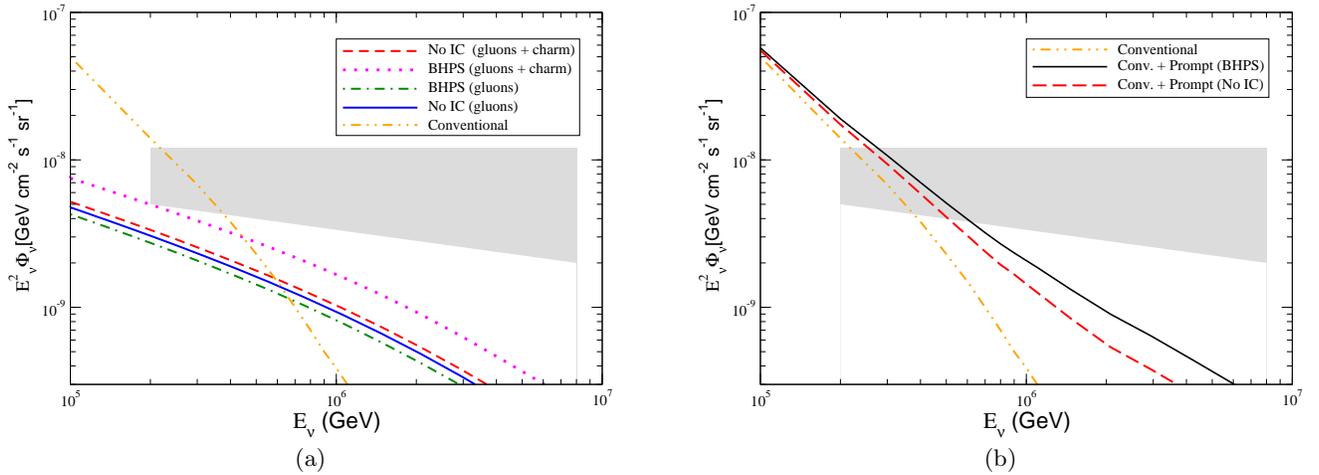

 \begin{center}
 \begin{tabular}{ccc}
 \includegraphics[width=0.45\textwidth]{fluxE2.eps} & \,\,\,\,\,\hspace{0.5cm}\,&
 \includegraphics[width=0.45\textwidth]{fluxE2_soma.eps} \\
 (a) & \,\,& (b)
 \end{tabular}
  \end{center}
 \caption{Energy dependence of the prompt neutrino flux, normalized by a factor $E_{\nu}^2$, obtained considering the BPL model for the primary nucleon spectrum. 
(a) Comparison of our results  with  the predictions for the conventional neutrino 
flux derived in Ref. \cite{Honda:2006qj} and the astrophysical neutrino flux obtained in Ref. \cite{Aartsen:2016xlq}, which is represented by the cyan band. (b) Comparison between the predictions for the  atmospheric (conventional + prompt) neutrino fluxes and the astrophysical one.}
  \label{Fig:fluxe2}
\end{figure}

\begin{figure}[t]
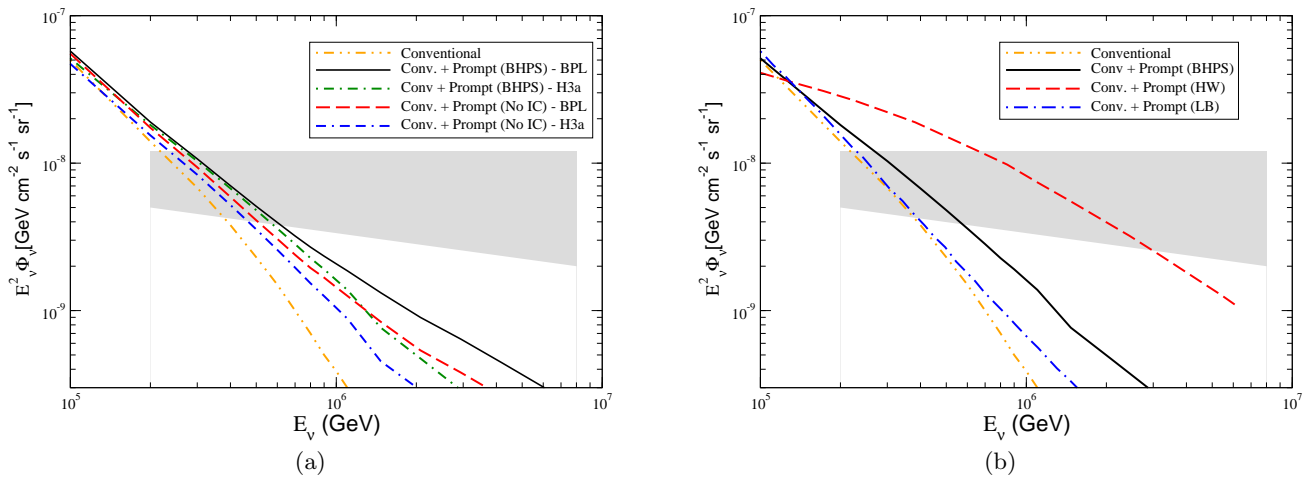

 \begin{center}
 \begin{tabular}{ccc}
 \includegraphics[width=0.45\textwidth]{fluxE2_soma_gaisser3.eps} & \,\,\,\,\,\hspace{0.5cm}\,&
 \includegraphics[width=0.45\textwidth]{fluxE2_soma_gaisser2_Comp2.eps} 
 \\
 (a) & \,\,& (b)
  \end{tabular}
  \end{center}
 \caption{(a) Comparison between the BPL and H3a predictions for the energy dependence of the atmospheric (conventional + prompt) neutrino fluxes, normalized by a factor $E_{\nu}^2$. (b) Comparison between the BHPS predictions for the atmospheric (conventional + prompt) neutrino flux with the results derived by Halzen and Willie (HW)  \cite{halzen} and by Laha and Brodsky (LB) \cite{laha}. Results derived using the H3a spectrum. }
  \label{Fig:fluxe2h3a}
\end{figure}

Using its six-year experimental data, the ICECUBE Collaboration has  recently  \cite{Aartsen:2016xlq} estimated  the astrophysical neutrino 
spectrum with the help of an unbroken power - law model. In order to compare our predictions 
with the   neutrino flux derived in Ref. \cite{Aartsen:2016xlq}, in Fig. \ref{Fig:fluxe2} we present 
our results for the neutrino flux, normalized by a factor $E_{\nu}^2$, obtained considering the BPL model for the primary nucleon spectrum.  The results from Ref. \cite{Aartsen:2016xlq} 
are represented by the gray band. For the sake of comparison, we also present the predictions for the conventional atmospheric 
neutrino flux derived in Ref. \cite{Honda:2006qj}. In Fig. \ref{Fig:fluxe2} (a) we compare our predictions 
for the prompt neutrino flux, considering different models and channels, with the conventional and astrophysical 
predictions. When the IC component and/or the charm channel is disregarded, the prompt contribution 
dominates for $E_{\nu} \gtrsim 6 \times 10^5$ GeV. On the other hand, if the IC component and the charm channel are 
taken into account, this transition occurs at $E_{\nu} \gtrsim 4.5 \times 10^5$ GeV. 
The most important aspect is that, although the astrophysical neutrino flux is dominant at large energies, the magnitude 
of the background associated to the prompt flux is strongly dependent on the presence or absence of the intrinsic component 
in the proton wave function, in agreement with the result obtained in Ref. \cite{laha}. This dependence  is more visible 
in the right panel of Fig. \ref{Fig:fluxe2}, where  we present  our predictions for the 
sum of the conventional and prompt fluxes and compare them with the ICECUBE results for the astrophysical neutrino flux. 
We observe that the inclusion of the IC component leads to an enhancement of the atmospheric (conventional + prompt) neutrino flux 
of  $\approx 45 \%$ at $E_{\nu} = 10^6$ GeV and of  $\approx 200 \%$ at $E_{\nu} = 3.5 \times 10^6$ GeV. Such large values 
are in agreement with the expectation discussed in Ref. \cite{prdnos}.

In Fig. \ref{Fig:fluxe2h3a} (a) we compare the BPL predictions with those obtained using the H3a model for the primary nucleon spectrum. As expected from our previous analysis, the atmospheric (conventional + prompt) neutrino flux derived using the H3a model is smaller the BPL one at large neutrino energies and they are similar to $E_{\nu} \approx 10^5$ GeV.  Finally, in Fig. \ref{Fig:fluxe2h3a} (b)  we compare the BHPS predictions, derived using the H3a spectrum, with the corresponding results obtained in Refs. \cite{halzen,laha}, considering different assumptions for the intrinsic charm component. In particular, in Ref. \cite{halzen} the authors have derived an upper limit for the 	neutrino spectrum assuming a maximum value for the contribution associated to the hadronization of the spectator charm with the valence quarks of projectile. This upper limit is denoted by HW in the figure. On the other, in Ref. \cite{laha} (denoted LB hereafter) the normalization of the intrinsic charm contribution for the $x_F$ - distribution is constrained by the data at low energies presented by the ISR experiments and LEBC - MPS Collaboration, with the energy dependence being that of the inelastic $pp$ cross section (See Ref. \cite{laha} for details). We have that our predictions are below from the BW one at large energies, with the difference between the results increasing with the neutrino energy. On the other hand, we predict a larger atmospheric neutrino flux than the LB model in the range probed by the IceCube. We believe that this difference is mainly associated to the energy dependence of the intrinsic contribution, which in our calculations is steeper than the energy dependence of the inelastic $pp$ cross section. As a consequence, our contribution of the charm initiated process increases faster with the center - of - mass energy than the LB one. 
The discrimination between these different predictions can, in principle, be feasible by future IceCube measurements dedicated to contrain the magnitude of the prompt contribution.

At this point, some comments are in order. In our calculations we  have  considered only  the BHPS model of intrinsic 
charm, in which the amount of IC is maximal.  From the results presented in Ref. \cite{prdnos}, we can expect that if the Meson 
Cloud  model is considered, we will get similar results for the enhancement associated to the IC component. However, if the amount of 
momentum  carried by the IC component is reduced, this will change our predictions for the prompt neutrino flux. 
Therefore, our predictions should be considered as an upper limit for the impact of the IC component. Another important aspect 
that should be emphasized is that we have used the color dipole approach to estimate the prompt neutrino flux. As discussed 
in detail in Refs. \cite{enb_stasto2,vicantoni}, this model implies a $x_F$ - distribution that is larger at intermediate 
values of $x_F$ than those obtained with the collinear and with the $k_T$ - factorization formalisms. As explained in  Refs. 
\cite{enb_stasto2,vicantoni}, such behavior is somewhat unexpected as this approach includes saturation effects that 
should lead rather to a reduction of the cross section compared to the collinear and the $k_T$ factorization approaches. The 
explanation of this difference  is still an open question and theme of debate (For a more detailed discussion see 
Ref. \cite{vicantoni}). Consequently, the color dipole predictions should be considered as upper bounds for the 
prompt neutrino flux.  The mentioned aspects imply that our 
predictions should be considered as upper bounds. However, we believe that our results strongly indicate 
that the inclusion of the charm channel in the color dipole approach for the heavy quark production is 
important to obtain realistic predictions and that the prompt neutrino flux is sensitive to the presence 
or absence of an IC component in the hadron wave function.

\section{Summary}
\label{conc}

 A complete knowledge of the partonic structure of  hadrons 
is fundamental to make predictions for the Standard Model and  beyond Standard Model processes observed  at 
hadron colliders. In particular, the heavy quark component of the 
proton  has a direct impact on the calculation of the prompt 
atmospheric neutrino flux, which is a background to the astrophysical 
neutrino flux  measured by the ICECUBE  Collaboration. 
One important (and not yet known) quantity for
heavy quark production is the amount of the intrinsic component in the hadron wave function. It carries a large fraction of 
the hadron momentum and, consequently, is expected to modify the cross sections and associated distributions of the produced  
heavy quarks and heavy mesons at forward rapidities. In this paper we have estimated the impact of the intrinsic charm component 
on the prompt neutrino flux using the color dipole approach to compute 
the charm $x_F$ - distribution. Following Ref. \cite{prdnos},  
we have generalized previous color dipole calculations by taking into account  the contribution of processes initiated by charm 
quarks. Moreover, differently from previous studies of IC effects, 
we have used in our calculations the parton 
distributions derived from the global analysis of a large set of experimental data, with evolution described by the DGLAP equations. 
Additionally, we have used as input in our calculations a model for the dipole - proton scattering amplitude that describes very well 
particle production at forward rapidities and LHC energies.  Our results indicate that the inclusion of the channel initiated by 
charm quarks has a strong effect on the prompt neutrino flux. In particular, if an IC component is present in the hadron wave 
function, our results indicate that the flux is enhanced by a factor 2 at large neutrino energies. Furthermore, we find that the 
astrophysical neutrino flux becomes dominant at $E_{\nu} \approx 10^6$ GeV. However, the magnitude of the background is strongly 
sensitive to the description of the prompt neutrino flux. Consequently, in order to disentangle the magnitude of the astrophysical 
contribution to the neutrino flux, it is mandatory to have a better theoretical and experimental control of the prompt neutrino flux.

\section*{Acknowledgments}
VPG thanks Rikard Enberg for provide the results obtained in Refs. \cite{ers,enb_stasto2} and Diego Rossi Gratieri by useful discussions regarding the conventional neutrino flux. A.V.G. gratefully acknowledges the Brazilian 
Funding Agency FAPESP for financial support (contract:2017/14974-8). This work was partially financed by the Brazilian funding agencies CAPES, CNPq,  FAPESP, FAPERGS and  INCT-FNA (process number 
464898/2014-5). Finally, the authors would like to thank the referee for
the careful reading of the manuscript and valuable comments that contributed to the improvement of the paper.

\end{document}